\documentclass[twocolumn,prl,showpacs]{revtex4}

\usepackage{dcolumn}
\usepackage{amsfonts}
\usepackage{amsmath}
\usepackage{amssymb}
\usepackage{bm}

\newif\ifpdf
\ifx\pdfoutput\undefined
\pdffalse 
\else
\pdfoutput=1 
\pdftrue \fi  \ifpdf
\usepackage[pdftex]{graphicx}
\else
\usepackage{graphicx}
\fi

\begin{document}

\ifpdf \DeclareGraphicsExtensions{.jpg,.pdf,.tif} \else
\DeclareGraphicsExtensions{.eps,.jpg} \fi

\newcommand{\brm}[1]{\bm{{\rm #1}}}
\newcommand{\tens}[1]{\underline{\underline{#1}}}
\newcommand{\mm}{\overset{\leftrightarrow}{m}}
\newcommand{\xv}{\bm{{\rm x}}}
\newcommand{\Rv}{\bm{{\rm R}}}
\newcommand{\uv}{\bm{{\rm u}}}
\newcommand{\nv}{\bm{{\rm n}}}
\newcommand{\Nv}{\bm{{\rm N}}}
\newcommand{\ev}{\bm{{\rm e}}}

\title{Dynamics, dynamic soft elasticity and rheology of smectic-$C$ elastomers}

\author{Olaf Stenull}
\affiliation{Fachbereich Physik, Universit\"at Duisburg-Essen,
Campus Essen, 45117 Essen, Germany }

\author{T. C. Lubensky}
\affiliation{Department of Physics and Astronomy, University of
Pennsylvania, Philadelphia, PA 19104, USA }

\vspace{10mm}
\date{\today}

\begin{abstract}
\noindent We present a theory for the low-frequency,
long-wavelength dynamics of soft smectic-$C$ elastomers with
locked-in smectic layers. Our theory, which goes beyond pure
hydrodynamics, predicts a dynamic soft elasticity of these
elastomers and allows us to calculate the storage and loss moduli
relevant for rheology experiments as well as the mode structure.
\end{abstract}


\maketitle

\noindent Smectic elastomers~\cite{WarnerTer2003} are rubbery
materials that have the macroscopic symmetry properties of smectic
liquid crystals~\cite{deGennesProst93_Chandrasekhar92}. They are sure to have intriguing properties,
some of which have already been studied experimentally and/or
theoretically~\cite{WarnerTer2003}. Very recently, seminal progress has been made on smectic-$C$ (Sm$C$) elastomers forming spontaneously from a smectic-$A$ (Sm$A$) phase upon cooling. Hiraoka {\em et al}.~\cite{hiraoka&CO_2005} for the first time produced a monodomain sample of such a material and carried out experiments demonstrating its spontaneous and reversible deformation in a heating an cooling process. Also very recently, it was discovered theoretically that such a material exhibits the fascinating phenomenon of soft elasticity~\cite{stenull_lubensky_SmCsoftness}, i.e., certain elastic moduli vanish as a consequence of the spontaneous symmetry breaking wherefore strains along specific symmetry direction cost no elastic energy and thus cause no restoring forces.

On one hand, due to the aforementioned experimental advances, dynamical experiments on soft Sm$C$ elastomers, such as rheology experiments of storage and loss moduli or Brillouin scattering measurements of sound velocities, seem within reach. On the other hand, there exists, to our knowledge, no dynamical theory that could be helpful in interpreting these kinds of experiments. Here we present a theory for the low-frequency, long-wavelength dynamics of soft Sm$C$ elastomers
with locked-in layers that goes beyond pure hydrodynamics. As in standard elastic media and nematic
elastomers~\cite{stenull_lubensky_2004} a purely hydrodynamical
theory of Sm$C$ elastomers involves
only a displacement field $\uv$ and not the Frank director $\nv$,
which relaxes to the local strain in a nonhydrodynamic time
$\tau_n$. We go beyond hydrodynamics, by including $\nv$ in our theory, because dynamical experiments, like rheology measurements, typically probe a wide range of frequencies that extends from hydrodynamic regime to frequencies well above it.

Smectic elastomers are, like any elastomers, permanently
crosslinked amorphous solids whose static elasticity is most
easily described in Lagrangian coordinates in which $\brm{x}$
labels a mass point in the undeformed (reference) material and
$\brm{R}(\brm{x})= \brm{x} + \brm{u}(\brm{x})$, where
$\brm{u}(\brm{x})$ is the displacement variable, labels the
position of the mass point $\brm{x}$ in the deformed (target)
material. Lagrangian elastic energies are formulated in terms of
the strain tensor $\tens{u}$ which, in its linearized form, has
the components $u_{ij} = \textstyle{\frac{1}{2}} (\eta_{ij} +
\eta_{ji})$, where $\eta_{ij} = \partial_j u_i$ are the components
of the displacement gradient tensor $\tens{\eta}$.

The elastic energy density $f$ of the Sm$C$ elastomers of interest
here can be divided into two
parts~\cite{stenull_lubensky_SmCsoftness}
\begin{align}
\label{formOfF}
f = f_{\brm{u}} +  f_{\brm{u}, \brm{n}} ,
\end{align}
where $f_{\brm{u}}$ depends only on $\tens{u}$ and $f_{\brm{u},
\brm{n}}$ describes the dependence of $f$ on the Frank director
$\brm{n}$ including its coupling to the displacement variable. In
the following we choose the coordinate system so that the $z$-axis
is parallel to the director of the initial Sm$A$ phase and the
$x$-axis is parallel to the direction of tilt in the resulting
Sm$C$ phase so that the equilibrium director characterizing the
undeformed Sm$C$ phase is of the form $\brm{n}^0 = (c, 0,
\sqrt{1-c^2})$ with $c$ being the order parameter of the
transition. With these conventions, $f_{\brm{u}}$ can be written
in the same form as the elastic energy density of conventional
monoclinic solids~\cite{monoclinic},
\begin{align}
\label{fu}
&f_{\brm{u}} =  \textstyle{\frac{1}{2}} \, C_{xyxy} \, u_{xy}^2 +
C_{xyzy} \, u_{xy} u_{zy} + \textstyle{\frac{1}{2}} \, C_{zyzy} \,
u_{zy}^2
\nonumber \\
& +\textstyle{\frac{1}{2}} \, C_{zzzz} \, u_{zz}^2  +
\textstyle{\frac{1}{2}} \, C_{xxxx} \, u_{xx}^2+
\textstyle{\frac{1}{2}} \, C_{yyyy} \, u_{yy}^2 +
\textstyle{\frac{1}{2}} \, C_{xzxz} \, u_{xz}^2
\nonumber \\
&+ C_{zzxx} \, u_{zz} u_{xx} + C_{zzyy} \, u_{zz} u_{yy}    +
C_{xxyy} \, u_{xx} u_{yy}
\nonumber \\
& + C_{xxxz} \, u_{xx} u_{xz}+ C_{yyxz} \, u_{yy} u_{xz} +
C_{zzxz} \, u_{zz} u_{xz} \, ,
\end{align}
but with constraints relating the three elastic constants in the
first row.  These latter constants can be expressed in terms of an
overall elastic constant $\bar{C}$ and an angle $\theta$, which
depends on the order parameter $c$, as $C_{xyxy} = \bar{C} \cos^2
\theta$, $C_{xyzy} = \bar{C} \cos \theta \sin \theta$, and
$C_{zyzy} = \bar{C} \sin^2 \theta$. Neglecting contributions from
the Frank energy, $f_{\brm{u}, \brm{n}}$ can be stated as
\begin{align}
\label{fun}
&f_{\brm{u}, \brm{n}} =  \textstyle{\frac{1}{2}} \, \Delta \, [
Q_y + \alpha  \, u_{xy} + \beta  \, u_{yx} ]^2
\end{align}
where $\Delta$ is a coupling constant and where $\alpha$ and
$\beta$ are dimensionless parameters. The variable $Q_y$ stands
for $Q_y =  n_y - c \, \eta_{Ayx} -  \sqrt{1-c^2} \, \eta_{Ayz}$,
where $\eta_{Ayx}$ and $\eta_{Ayz}$ are components of the
antisymmetric part of $\tens{\eta}$.

Equations~(\ref{formOfF}) to (\ref{fun}) imply that Sm$C$
elastomers are soft under static conditions. Imagine that $Q_y$
has relaxed locally to $Q_y  = -\alpha  \, u_{xy} -  \beta  \,
u_{yx}$ so that $f$ is effectively reduced to $f_{\brm{u}}$. Then,
due to the above relations among the elastic constants,
deformations characterized by Fourier transformed displacements
$\brm{u}\parallel \hat{e}_y$ and wavevectors $\brm{q} \parallel
\hat{e}_2 = (-\sin \theta, 0, \cos \theta)$ or, alternatively,
$\brm{u}
\parallel \hat{e}_2$ and $\brm{q} \parallel \hat{e}_y$ cost no
elastic energy and hence cause no restoring forces.
The effects of soft elasticity are more evident in a coordinate
systems rotated though $\theta$ about the $y$-axis in which
$\hat{e}_{x^\prime} = (\cos \theta, 0, \sin \theta)$,
$\hat{e}_{y^\prime} = \hat{e}_y$, and $\hat{e}_{z^\prime} =
\hat{e}_2$.  In this system, $C_{x^\prime y^\prime z^\prime
y^\prime}$ and $C_{z^\prime y^\prime z^\prime y^\prime}$ vanish
impyling there is no energy cost for shears $u_{z^\prime
y^\prime}$ in the $z^\prime y^\prime$-plane.  If the elastomer is
crosslinked in the Sm$C$ phase, these moduli become nonzero. Here,
we will not consider such semi-soft Sm$C$ elastomers.

Now, let us formulate our dynamical
theory. Dynamical equations for $\uv$ and $\nv$ can be derived
using standard Poisson-bracket
approaches~\cite{forster&Co_71_Forster1983}, with the
result~\cite{stenull_lubensky_2004}
\begin{subequations}
\label{genStruct}
\begin{align}
\label{EOMa} \dot{n}_i &= \lambda_{ijk} \, \partial_j
\dot{u}_k - \Gamma \, \frac{\delta \mathcal{H}}{\delta n_i}\, ,
\\
\label{EOMb} \rho \ddot{u}_i &= \lambda_{kji} \, \partial_j
\frac{\delta \mathcal{H}}{\delta n_k}  - \frac{\delta
\mathcal{H}}{\delta u_i} + \nu_{ijkl}\, \partial_j
\partial_l \dot{u}_k \, ,
\end{align}
\end{subequations}
where $\mathcal{H}$ is the elastic energy of the system,
$\nu_{ijkl}$ is the viscosity tensor, $\Gamma$ is a dissipative
coefficient with dimensions of an inverse viscosity, and
\begin{eqnarray}
\lambda_{ijk} = \textstyle{\frac{1}{2}} \, \lambda \, \left(
\delta_{ij}^T \, n_k  + \delta_{ik}^T \,n_k
\right)+\textstyle{\frac{1}{2}} \, \left( \delta_{ij}^T \, n_k  -
\delta_{ik}^T \,n_k \right) ,
\end{eqnarray}
where $\delta_{ij}^T = \delta_{ij} - n_i n_j$. As they stand,
Eqs.~(\ref{genStruct}) are valid for any liquid crystal elastomer
with a defined director, like e.g., nematic, Sm$A$ and Sm$C$
elastomers. To describe Sm$C$ elastomers we have have to specify
$\mathcal{H}$ and $\nu_{ijkl}$ accordingly. From the above it is
clear that $\mathcal{H} = \int d^3 x f$ with $f$ as given in
Eq.~(\ref{formOfF}). The viscosity tensor entering here is that of
a monoclinic system. It has 13 independent components and it can be
parameterized, as we do, so that the entropy production density $T
\dot{s}$ takes on the same form as Eq.~(\ref{fu}) with the elastic
constants $C_{ijkl}$ replaced by viscosities $\nu_{ijkl}$ and with
$u_{ij}$ replaced by $\dot{u}_{ij}$. The $\nu_{ijkl}$ depend on the order parameter $c$. $\nu_{xxxz}$, $\nu_{yyxz}$, and $\nu_{zzxz}$ vanish at the Sm$C$ to Sm$A$ transition and are therefore expected being smaller than the remaining viscosities for $c$ small~\cite{osipov&Co_1995}.

A smectic elastomer is characterized in general by relaxation
times associated with director relaxation and with other modes,
which we will simply refer to as elastomer modes. For frequencies
$\omega$ such that $\omega \tau_E \ll1$, where $\tau_E$ is the
longest elastomer time, the viscosities and $\Gamma$ are
practically frequency independent. When $\omega \tau_E \geq 1$,
however, the viscosities $\nu_{ijkl}$ and $\Gamma$ develop a
non-trivial frequency dependence. In the following we will
consider in detail only the case $\tau_n \gg \tau_E$ and $\omega
\tau_E \ll1$.

As mentioned above, Eq.~(\ref{fun}) omits contributions from the
Frank elastic energy for director distortions, which are higher
order in derivatives than those arising from network elasticity.
Without the Frank energy, our dynamical theory misses diffusive
modes along certain symmetry directions where sound velocities
vanish. Including the Frank energy, on the other hand, makes the
equations of motion considerably more complicated. To keep our
presentation as simple as possible, we will therefore, for the
most part,  exclude the Frank energy. When it comes to stating
results for the aforementioned diffusive modes, however, we will
include Frank contributions in order to present complete results.

From Eq.~(\ref{EOMa}) we can derive an equation of motion for
$Q_y$. In frequency space, this equation can be written as
\begin{align}
\label{QyFourier}
Q_y &= - \alpha \, \frac{1 + i \omega \tau_3}{1 - i \omega \tau_1}
\,  u_{yx} - \beta \, \frac{1 + i \omega \tau_2}{1 - i \omega
\tau_1} \,  u_{yz} \,
\end{align}
where we have introduced the relaxation times $\tau_1 = 1/( \Gamma
\Delta)$, $\tau_2 = \lambda \sqrt{1-c^2}/(\Gamma \Delta \beta)$,
$\tau_3 = \lambda c/( \Gamma \Delta \alpha)$. As we will see
further below, our dynamical equations predict nonhydrodynamic
modes with a decay time (``mass") $\tau_1$ which implies $\tau_1 =
\tau_n$.

With help of Eqs.~(\ref{QyFourier}) and (\ref{EOMb}) we derive
effective equations of motions in terms of the displacements only
which can be cast as
\begin{align}
\label{EOMdense}
\rho \omega^2 u_i = - \partial_j \sigma_{ij} (\omega)
\end{align}
with a symmetric stress tensor $\tens{\sigma}$ given by
\begin{subequations}
\label{effectiveEOMs}
\begin{align}
\sigma_{\mu} (\omega) &= C_{\mu \nu} (\omega) \, u_{\nu} + C_{\mu
xz} (\omega) \, u_{xz} \, ,
\\
\sigma_{xz} (\omega) &= \textstyle{\frac{1}{2}} \,C_{xz \nu}
(\omega) \, u_{\nu} + \textstyle{\frac{1}{2}} \, C_{xzxz} (\omega)
\, u_{xz} \, ,
\\
\sigma_{\xi} (\omega) &= \textstyle{\frac{1}{2}} \, C_{\xi \chi}^R
(\omega) \, u_{\chi}  \, ,
\end{align}
\end{subequations}
where we use a compact notation with indices $\mu$, $\nu$ running
over $xx$, $yy$ and $zz$ and  indices $\xi$, $\chi$ running over
$xy$ and $zy$. $C_{ijkl} (\omega)$ with no superscript $R$ stands
for $C_{ijkl} (\omega) = C_{ijkl} - i \omega \nu_{ijkl}$. The
superscript $R$ indicates that certain elastic moduli are
renormalized by the relaxation of the director. These are
\begin{align}
\label{renormalizedModuli}
C_{\xi \chi}^R (\omega) &=  C_{\xi \chi} - i \omega  \, \nu_{\xi
\chi}  - \frac{i\omega \tau_1}{1 - i\omega \tau_1} \,  \Delta
A_{\xi \chi}
\nonumber \\
&= C_{\xi \chi} - i \omega  \, \nu_{\xi \chi}^R + O(\omega^2),
\end{align}
with renormalized viscosities $\nu_{\xi \chi}^R = \nu_{\xi \chi} +
\Gamma^{-1} A_{\xi \chi}$, and where $A_{xyxy} = \alpha^2 (1+
\tau_3/\tau_1)^2$, $A_{xyzy} = \alpha \beta (1+ \tau_3/\tau_1)(1+
\tau_2/\tau_1) $ and $A_{zyzy} = \beta^2 (1+ \tau_2/\tau_1)^2$.

\begin{figure}
\includegraphics[width=5.5cm]{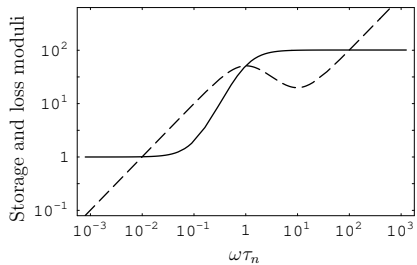}
\caption[]{\label{LossStorageModuli}Log-log plot of the reduced storage
and loss moduli $C_{\xi \chi}^\prime (\omega)/C_{\xi \chi}$ (solid
line) and $C_{\xi \chi}^{\prime \prime}(\omega)/C_{\xi \chi}^R $
(dashed line) versus the reduced frequency $\omega \tau_n$ as
given respectively by the real and negative imaginary parts of
Eq.~(\ref{renormalizedModuli}) for $\tau_n \gg \tau_E$. For the
purpose of illustration we have set, by and large arbitrarily,
$\nu_{\xi \chi} /(\tau_n C_{\xi \chi}) = 1$ and $\Delta A_{\xi
\chi} /C_{\xi \chi} = 10^2$.}
\end{figure}

The frequency dependence of the elastic moduli in
Eq.~(\ref{effectiveEOMs}) can be determined, in principle, by
rheology measurements of the corresponding storage and loss
moduli. The unrenormalized moduli $C_{ijkl} (\omega)$ lead to
conventional storage and loss moduli $C_{ijkl}^\prime = C_{ijkl}$
and $C_{ijkl}^{\prime\prime} = \omega \nu_{ijkl}$ that are, as in
conventional rubbers, respectively constant and proportional to
$\omega$ at low frequencies. The renormalized moduli, on the other
hand, have the potential for much more interesting rheology
behavior. One consequence of Eq.~(\ref{renormalizedModuli}) is
that Sm$C$ elastomers could exhibit so-called dynamic soft
elasticity~\cite{terentjev&Co_NEhydrodyn}. To highlight this
phenomenon, let us switch briefly to the rotated reference space
coordinates $x^\prime, y^\prime, z^\prime$. Then,
Eq.~(\ref{renormalizedModuli}) implies that
\begin{align}
\label{renormalizedModuliPrime}
C_{y^\prime z^\prime y^\prime z^\prime}^R (\omega) &=   - i \omega
\, \nu_{y^\prime z^\prime y^\prime z^\prime}  - \frac{i\omega
\tau_1}{1 - i\omega \tau_1} \,  \Delta A_{y^\prime z^\prime
y^\prime z^\prime}
\end{align}
with $\nu_{y^\prime z^\prime y^\prime z^\prime} = \sin^2 \theta
\nu_{xyxy} - \sin 2 \theta \nu_{xyyz} + \cos^2 \theta \nu_{yzyz}$
and an analogous expression for $A_{y^\prime z^\prime y^\prime
z^\prime}$. $C_{x^\prime y^\prime y^\prime z^\prime}^R (\omega)$
is of the same form as Eq.~(\ref{renormalizedModuliPrime}). Thus,
$C_{y^\prime z^\prime y^\prime z^\prime}^R (\omega)$ and
$C_{x^\prime y^\prime y^\prime z^\prime}^R (\omega)$ vanish in the
limit $\omega \to 0$ where we recover true soft elasticity. At
non-vanishing frequency the system cannot be ideally soft but it
can be nearly so for $\omega$ small. This type of behavior was
first predicted for nematic elastomers, where it has been termed
dynamic soft elasticity~\cite{terentjev&Co_NEhydrodyn}. The
storage moduli for $u_{xy}$ and $u_{yz}$ strains,
Eq.~(\ref{renormalizedModuli}), in the original coordinate system
are nonzero at zero frequency. Their behavior for $\omega >0$,
like that of semi-soft nematic
elastomers~\cite{terentjev&Co_NEhydrodyn,stenull_lubensky_2004},
depends on $\tau_n/ \tau_E$.  If $\tau_n \gg \tau_E$, the storage
moduli exhibit a step and the corresponding loss moduli an
associated peak at $\omega \tau_n \sim 1$ as shown in
Fig.~\ref{LossStorageModuli}; if $\tau_n \approx \tau_E$, or
$\tau_n < \tau_E$, this is not the case. The storage moduli
$C_{y^\prime z^\prime y^\prime z^\prime}^\prime (\omega)$ and
$C_{x^\prime y^\prime y^\prime z^\prime}^\prime (\omega)$ in the
rotated frame are zero at zero frequency.  In a semi-soft Sm$C$,
however, they will exhibit behavior similar to that
Fig.~\ref{LossStorageModuli} for $\tau_n \gg \tau_E$. In nematic
elastomers, there is still some controversy~\cite{controversy}
about whether the $\tau_n \gg \tau_E$ regime has actually been
observed in experiments. It would be interesting to see if it
might exist in Sm$C$ elastomers, where $\tau_E$ might be shorter
than it is in nematics because of the smectic layers.

\begin{figure}
\includegraphics[width=7.0cm]{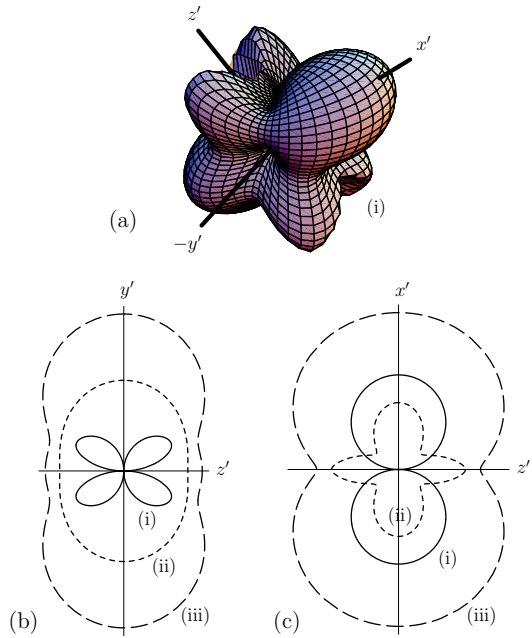}
\caption[]{\label{fig:soundVelocities}Schematic plots of sound
velocities: (a) spherical plot of mode pair (i) only; (b) and (c)
polar plots in the $y^\prime z^\prime$- and $x^\prime z^\prime$-
planes, respectively, of all 3 sound-mode pairs.}
\end{figure}

To assess the mode structure of Sm$C$ elastomers, we start with an
analysis of propagating sound modes in the dissipationless limit.
The sound modes have frequencies $\omega (\brm{q}) = C (\vartheta,
\varphi) q $, where $q = |\brm{q}|$ and where $\vartheta$ and
$\varphi$ are the azimuthal and polar angles of $\brm{q}$ in
spherical coordinates. Their sound velocities $C (\vartheta,
\varphi)$, as calculated from Eq.~(\ref{EOMdense}) with the
viscosities $\nu_{ijkl}$ set to zero, are depicted in
Fig.~\ref{fig:soundVelocities}. There are 3 pairs of sound modes.
One of these pairs (i) is associated with the soft deformations
discussed above. Its velocity vanishes for $\brm{q}$ along
$\hat{e}_y$ and $\hat{e}_2$ so that when viewed in the $y^\prime
z^\prime$-plane it has a clover leave like shape. The remaining 2
pairs are associated with non-soft deformations. In the
incompressible limit, these pairs become purely transverse (ii)
and longitudinal (iii), respectively. In the $y^\prime z^\prime$
and $x^\prime z^\prime$-planes, their velocities are non-vanishing
in all directions. Note that, since the velocity of pair (i)
vanishes in directions where the velocities of the other modes
remain finite, Sm$C$ elastomers are, like nematic
elasomers~\cite{terentjev&Co_NEhydrodyn}, potential candidates for
applications in acoustic polarizers.

Having found the general sound-mode structure in the
nondissipative limit, we now turn to the full mode structure in
the incompressible limit. In the softness-related symmetry
directions the modes are of the following types: (i)
non-hydrodynamic modes with frequencies
\begin{align}
\label{wm}
\omega_{m} = - i \tau_1^{-1} + i D_m \, q^2
\end{align}
with a zero-$q$ decay time $\tau_1$ and diffusion constants $D_m$,
(ii) propagating modes with frequencies
\begin{align}
\label{wp}
\omega_{p} = \pm C\,  q -  i D_p \, q^2
\end{align}
with sound velocities $C$ and diffusion constants $D_p$, and (iii)
diffusive modes with frequencies
\begin{align}
\label{wd}
\omega_{d} = -  i D_d \, q^2 \pm \sqrt{- (D_d \, q^2)^2 + B \,
q^4}
\end{align}
with diffusion constants $D_d$ and bending terms $B$ that are
missed if the Frank energy is neglected. For $B/D_d^2 \ll 1$ the
diffusive modes split up into slow and fast modes
\begin{align}
\label{wsf}
\omega_{s} = -  i B/(2D_d) \, q^2 \, , \qquad  \omega_{f} = -  i
2D_d \, q^2 \, .
\end{align}
Specifics of the sound velocities, the diffusion constants and the
bending terms are given in the following.

First, let us consider the case that $\brm{q}$ lies in the
$xz$-plane. In this case the equation of motion for $u_y$
decouples from the equations of motion for $u_x$ and $u_z$. This
equation produces a set  of transverse modes with $\brm{u}
\parallel \hat{e}_y$. There is a non-hydrodynamic mode with
\begin{align}
\label{wym}
D_{m,y}=  (4\rho)^{-1} \left[  \sqrt{\nu_{xyxy}^R - \nu_{xyxy}} \,
\hat{q}_x + \sqrt{\nu_{zyzy}^R - \nu_{zyzy}} \, \hat{q}_z
\right]^2 ,
\end{align}
where $\hat{q}_i = q_i/q$, and there are propagating modes with
\begin{subequations}
\begin{align}
C_{y}  &=  \sqrt{\bar{C}/(4\rho)}\,  | \cos \theta \, \hat{q}_x +
\sin \theta \, \hat{q}_z |=\sqrt{\bar{C}/(4\rho)}|\hat{q}_{x'}| \,
,
\\
D_{p,y}  &= (8\rho)^{-1} \left[  \nu_{xyxy}^R \hat{q}_x^2 + 2
\nu_{xyzy}^R \hat{q}_x \hat{q}_z + \nu_{zyzy}^R\hat{q}_z^2 \right]
. \,
\end{align}
\end{subequations}
In the soft direction, i.~e.\ for $\brm{q} \parallel
\hat{e}_{z^\prime}$, these propagating modes become diffusive with
$D_{d,y} = D_{p,y}$ and
\begin{align}
\label{wypmDiffusive}
B_y  =  \rho^{-1} \big[ \bar{K}_1 \hat{q}_x^4  + \bar{K}_2
\hat{q}_z^4 + \bar{K}_3 \hat{q}_x^2 \hat{q}_z^2 + 2 \bar{K}_4
\hat{q}_x^3 \hat{q}_z + 2 \bar{K}_5 \hat{q}_x \hat{q}_z^3 \big],
\end{align}
where the $\bar{K}$'s are bending moduli that are combinations of
the usual Frank elastic constants, the order parameter $c$ as well
as $\lambda$, $\alpha$, and $\beta$. The equations of motion for
$u_x$ and $u_z$ can be solved by decomposing $(u_x,u_z)$ into a
longitudinal part $u_l$ along $\brm{q}$ and a transversal part
$u_T$. In the incompressible limit $u_l$ vanishes. The equation of
motion for $u_T$ produces a pair of propagating modes with
\begin{subequations}
 \begin{align}
C_{T}  &=  \sqrt{1/\rho}\,  \big\{  [C_{xxxx} + C_{zzzz} - 2
C_{xxzz}] \hat{q}_x^2 \hat{q}_z^2
 \\
&+ [C_{xxxz}  -  C_{zzxz}] \hat{q}_x \hat{q}_z (\hat{q}_z^2
-\hat{q}_x^2) + \textstyle{\frac{1}{4}}  C_{xzxz}(\hat{q}_z^2
-\hat{q}_x^2)\big\}^{1/2} \nonumber
\\
D_{p,T} &=  (2\rho)^{-1} \big[  [\nu_{xxxx} + \nu_{zzzz} - 2
\nu_{xxzz}] \hat{q}_x^2 \hat{q}_z^2
 \\
&+ [\nu_{xxxz}  -  \nu_{zzxz}] \hat{q}_x \hat{q}_z (\hat{q}_z^2
-\hat{q}_x^2) + \textstyle{\frac{1}{4}}  \nu_{xzxz}(\hat{q}_z^2
-\hat{q}_x^2)\big] , \nonumber
\end{align}
\end{subequations}
and $\brm{u} \parallel \hat{e}_T$ where $\hat{e}_T = (\hat{q}_z,
0, -\hat{q}_x)$.

Finally, we turn to the case $\brm{q} \parallel \hat{e}_y$. There
is a pair of longitudinal propagating modes with $\brm{u}
\parallel \hat{e}_y$ that is suppressed in the incompressible
limit. There is a non-hydrodynamic mode with
\begin{align}
\label{wxzm}
D_{m,xz}=  (4\rho)^{-1} \left[  \nu_{xyxy}^R - \nu_{xyxy}+
\nu_{zyzy}^R - \nu_{zyzy}  \right]   ,
\end{align}
where $\brm{u}$ lays in the $xz$-plane with $u_x = \alpha (\tau_1
+\tau_3)/[\beta (\tau_1 +\tau_2)] \, u_z$. There is a pair of
elastically soft diffusive modes with polarization $\brm{u}
\parallel \hat{e}_{z^\prime}$ with
\begin{subequations}
\begin{align}
D_{d,z^\prime}  &=  (8\rho)^{-1}\,  \nu_{z^\prime y^\prime
z^\prime y^\prime}^R  \, ,
\\
B_{z^\prime}  &= \rho^{-1} \left[  \sin^2 \theta \,  \bar{K}_6 -
\sin 2 \theta \,\bar{K}_8   + \cos^2 \theta \, \bar{K}_7 \right] ,
\end{align}
\end{subequations}
where $ \nu_{z^\prime y^\prime z^\prime y^\prime}^R$ is the
renormalized version of $ \nu_{z^\prime y^\prime z^\prime
y^\prime}$ and where the $\bar{K}$'s are once more bending moduli
depending on the Frank constants, $c$, $\lambda$, $\alpha$, and
$\beta$. Finally, there is a pair of propagating modes polarized
along $\hat{e}_{x^\prime}$ with
\begin{subequations}
\begin{align}
C_{x^\prime}  &=  \sqrt{\bar{C}/(4\rho)}\, ,
\\
D_{p,x^\prime}  &=   (8\rho)^{-1} \,  \nu_{x^\prime y^\prime
x^\prime y^\prime }^R \,  ,
\end{align}
\end{subequations}
with $\nu_{x^\prime y^\prime x^\prime y^\prime }^R = \cos^2 \theta
\,  \nu_{xyxy}^R + \sin 2 \theta \, \nu_{xyzy}^R  + \sin^2 \theta
\, \nu_{zyzy}^R$.

In summary, we have presented a theory for the low-frequency,
long-wavelength dynamics of soft Sm$C$ elastomers. This theory
predicts that, at least in an idealized limit, Sm$C$ elastomers
possess the fascinating property of dynamic soft elasticity.
Though the equations of motion are complicated, the resulting mode
structure is with respect to the softness related symmetry
directions nicely symmetric and it has, when visualized, a certain
beauty. We have calculated various dynamical quantities, such as
storage and loss moduli and sound velocities that are, in
principle, accessible by experiments and we hope, that our theory
encourages such experimental work.

Support by the National Science Foundation under grant DMR 0404670 (TCL) is gratefully acknowledged.

\end{document}